# A Comparative Study of Hybrid Models in Health Misinformation Text Classification

Mkululi Sikosana
Manchester Metropolitan University, Department Of Computing And Mathematics
mkululi.sikosana@stu.mmu.ac.uk

Oluwaseun Ajao
Manchester Metropolitan University, Department Of Computing And Mathematics
s.ajao@mmu.ac.uk

Sean Maudsley-Barton
Manchester Metropolitan University, Department Of Computing And Mathematics
s.maudsley-barton@mmu.ac.uk

## ABSTRACT

This study evaluates the effectiveness of machine learning (ML) and deep learning (DL) models in detecting COVID-19-related misinformation on online social networks (OSNs), aiming to develop more effective tools for countering the spread of health misinformation during the pan-demic. The study trained and tested various ML classifiers (Naive Bayes, SVM, Random Forest, etc.), DL models (CNN, LSTM, hybrid CNN+LSTM), and pretrained language models (DistilBERT, RoBERTa) on the "COVID19-FNIR DATASET." These models were evaluated for accuracy, F1 score, recall, precision, and ROC, and used preprocessing techniques like stemming and lemmatization. The results showed SVM performed well, achieving a 94.41% F1-score. DL models with Word2Vec embeddings exceeded 98% in all performance metrics (accuracy, F1 score, recall, precision & ROC). The CNN+LSTM hybrid models also exceeded 98% across performance metrics, outperforming pretrained models like DistilBERT and RoBERTa. Our study concludes that DL and hybrid DL models are more effective than conventional ML algorithms for detecting COVID-19 misinformation on OSNs. The findings highlight the importance of advanced neural network approaches and large-scale pretraining in misinformation detection. Future research should optimize these models for various misinformation types and adapt to changing OSNs, aiding in combating health misinformation.

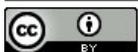

## 1 INTRODUCTION

The proliferation of health misinformation on online social networks (OSNs) has become a critical public health issue, especially during the COVID-19 pandemic. Estimates suggest that health misinformation on OSNs can range widely, from 0.2% to 28.8% [1]. With an estimated 3.6 to 4.7 billion individuals on a global level actively using OSNs [2–4] and a projected in-crease to 5.85 billion by 2027 [5, 6], the potential for exposure to misleading health information is vast and concerning.

The significance of this issue is highlighted by the severe consequences of health misinformation. For example, [7] reported substantial numbers of premature deaths (800 deaths) and 5 876 hospitalizations in Iran due to methanol consumption, a misguided COVID-19 remedy. These instances highlight the pressing need for reliable methods to identify and counteract health misinformation on OSNs.

However, the domain faces challenges, as traditional fact-checking methods struggle to keep pace with the volume of content on OSNs [8]. This gap in effective countermeasures has sparked controversy and highlighted the need for innovative solutions. The use of ML and DL models for misinformation detection has emerged as a promising approach, yet their effectiveness in the specific context of health misinformation on OSNs remains an area of active research and debate.

This research contributes to the field of COVID-19-health misinformation detection on OSNs by:

Replicating experiments: Initially replicating the experiments conducted by [19], providing a foundation for further investigation and validation of findings.

Evaluating ML and DL models: Assessing the efficacy of various ML algorithms and DL models in identifying COVID-19 health misinformation, thereby contributing to the body of knowledge on effective computational techniques for health misinformation detection.

Potential for tool development: Offering insights and potential pathways for the development of robust digital tools aimed at mitigating the adverse effects of health misinformation in the digital landscape, thus contributing to public health and information integrity.

Identifying effective models: The principal conclusion involves identifying specific models that demonstrate high accuracy and efficiency in health misinformation detection, marking a significant step toward developing more effective tools.

## 1.1 Related works

There is significant literature which contributes to the understanding of misinformation detection in online social networks (OSNs), particularly regarding COVID-19-related fake news. This literature points to the effectiveness of conventional machine learning algorithms, DL approaches, the need for context-based approaches, and the challenges in detecting misinformation that closely resembles the truth. In addition, this literature highlights the importance of tailored models for specific types of misinformation and offer insights into the evaluation and comparison of detection models. Subjecting the literature to [9] bottom-up thematic analytical approach returned the following themes, which provide a comprehensive overview of the current state of the art (SOTA) in health misinformation detection.

## 1.2 Deep learning approaches for misinformation detection

Current research has highlighted the critical role of DL models in detecting COVID-19 misinformation on OSNs, demonstrating their effectiveness with notable metrics. For ex-ample, [10] focused on the COVID-19 infodemic and used DL models such as LSTM, Bi-directional Long Short-Term Memory (Bi-LSTM) and Gated Recurrent Unit (GRU). Of note, their BiLSTM model achieved accuracy of 94% for short English sentences, 99% for long English sentences, and 82% for Chinese texts. These results highlight the model's adaptability and effectiveness in processing different lengths and languages of textual data.

Similarly, [11] developed an automated model using LSTM networks, integrating word embeddings such as CountVectoriser and TF-IDF. Their model demonstrated an accuracy of impressive 99.82%, thus surpassing traditional ML models and existing DL approaches. This indicates the potential of LSTM networks in effectively capturing the nuances of misinformation in textual content. On the other hand, [12] employed a CNN-based DL model for detecting COVID-19 fake news, achieving significant performance metrics such as a mean accuracy of 96.19%, a mean F1-score of 95%, and a high AUC-ROC of 98.5%. These results reflect the CNN model's capability in handling the complexity of fake news content. While both [11] and [12] demonstrate the efficacy of their respective models, a critical consideration is their applicability in real-world scenarios, where misinformation often involves evolving narratives and diverse formats. Although the studies by [11] and [12] achieved high accuracy, they could, however, benefit from a broader exploration of how these models perform in dynamic and heterogeneous OSN environments.

## 1.3 Multimodal and multichannel detection approaches

The evolution of misinformation into multimodal forms has necessitated the development of advanced detection methods. [13] highlight this shift in their survey, focusing on the transition to multimodal fake news detection in social media. While their study emphasizes the need for sophisticated detection techniques, it may fall short in providing detailed analysis of specific multimodal detection methods or empirical evidence of their effectiveness. This gap indicates a potential area for future research to explore and validate specific techniques for combating the increasingly complex landscape of misinformation.

In recent advancements within the domain of DL for health misinformation detection, the works of [14] and [15] stand out due to their novel contributions and significant performance metrics. These studies introduce innovative DL models that set new benchmarks in the accuracy and efficiency of detecting COVID-19-related misinformation across various datasets. In their work, [14] unveiled Vec4Cred, a model designed to evaluate the genuineness of health information, a particularly salient issue amidst the COVID-19 pandemic. The model's prowess was put to the test across various datasets, where it consistently demonstrated its superiority. Within the confines of the Microsoft Credibility Dataset, Vec4Cred achieved an accuracy of 88.25% alongside a 94.21% Area Under the Curve (AUC), metrics that testament its efficacy. The model further solidified its standing in the Medical Web Reliability Corpus, achieving near-perfect accuracy and AUC of 99.71%. Even in the diverse landscape of the CLEF eHealth 2020 Task-2 Dataset, Vec4Cred maintained robust performance with 82.56% accuracy and an 81.11% AUC. Despite these achievements, the exploration of Vec4Cred's versatility in addressing misinformation beyond the health domain remains a promising avenue for future research.

Similarly, [15] in their study, proposed a multichannel Convolutional Neural Network (CNN) model that distinguishes itself from its predecessors through its enhanced ability to detect COVID-19 fake news. This model's architecture, which leverages multiple channels to process information, has been shown to outclass single-channel CNNs and other contemporary models. The efficacy of this approach is highlighted by its performance metrics, boasting approximately 97% in accuracy, precision, recall, and F1-score across validation and test datasets. This remarkable achievement not only highlights the potential of multichannel approaches in improving misinformation detection, but also signals the need for further investigation into the model's adaptability and effectiveness across various misinformation scenarios.

The contributions of [14] and [15] represent significant strides in the fight against health misinformation, particularly in the context of the COVID-19. Thus, by pushing the boundaries of what DL models can achieve in terms of accuracy and efficiency, their work lays the groundwork for future advancements in the development of sophisticated tools for misinformation detection. As these models undergo further scrutiny and adaptation, their potential to impact the digital landscape remains vast, heralding a new era of reliability and trust in online health information.

## 1.4 Comparative analysis and future directions

Several studies have contributed to comparative analyses and offered insights into future research directions, often accompanied by relevant metrics. For example, [16] conducted a comprehensive review that explored the complex nature of fake news detection and highlighted the limitations of current AI approaches. While the study provides valuable insights into the challenges faced in this domain, it lacks a detailed analysis of practical implementation and real-world efficacy of the AI methodologies dis-cussed. This omission limits the study's applicability in providing actionable solutions for effectively countering fake news on OSNs. Furthermore, [16] do not offer specific metrics or empirical data to support their conclusions, which could have strengthened the study by demonstrating the practical implications of the highlighted limitations. Our research addresses this gap by providing a focused empirical evaluation of health misinformation detection models, specifically targeting the limitations outlined by [16]. Our study employs a range of performance metrics such as accuracy, and F1-score, offering an assessment of how these limitations impact model effectiveness across various misinformation scenarios. Thus, through integrating quantitative analysis with the theoretical concerns raised by [16] our work not only fills the identified literature gap, but also contributes practical insights towards enhancing health misinformation detection methodologies. This empirical foundation enriches the theoretical discourse, guiding future advancements in the development of robust detection models. Our findings emphasize the necessity of empirical validation in refining and validating theoretical models, paving the way for improved tools in combating digital misinformation.

[17] offer a comprehensive survey encompassing various attributes, features, and detection methods for fake news, covering news content, social context, and creators. The extensive theoretical analysis by [17] provides a valuable framework for understanding fake news detection. However, this study falls short in providing in-depth case studies or practical applications, which are crucial for applying theoretical insights to real-world scenarios. Furthermore, the rapidly changing nature of social media and fake news implies that some aspects of this survey might quickly become outdated, signaling a need for continuous updates and practical, contemporary examples to remain relevant and effective.

In reviewing the study by [18], we find valuable insights into the effective-ness of various fake news detection models. The study provides a comparative analysis of conventional ML classifiers and DL approaches, focusing on their generalization capabilities across different datasets. Notably, it reveals that conventional classifiers such as Naive Bayes and Random Forest often outperform DL models such as BERT and RoBERTa in terms of generalization, though no single model consistently emerges as the best. The study offers detailed metrics that are crucial for understanding the performance of these models in different contexts. For example, BERT achieved an accuracy of 98.7% on the ISOT Fake News dataset, 63.0% on LIAR, 96.0% on the "Fake News" dataset, 85.3% on FakeNewsNet, and 75.0% on the COVID-19 Fake News dataset. RoBERTa, on the other hand, scored 99.9% on ISOT, 67.4% on LIAR, and showed varied effectiveness on other datasets with 82.0% and 77.9% accuracy on COVID-19 datasets. These performance metrics highlight the different levels of effectiveness of these models under various conditions, providing valuable insights into their applicability and limitations.

It is evident from current studies in COVID-19 misinformation detection, that each study contributes unique insights, but also reveals certain limitations. For example, [19] provide a detailed examination of ML algorithms, with Random Forest and Stochastic Gradient Descent (SGD) showing promising results (91.6% accuracy and 92% F1 score for Random Forest; 91.5% accuracy and 92% F1 score for SGD). These metrics highlight the potential effectiveness of specific algorithms in health misinformation detection. However, the [19] study focuses primarily on model performance, lacking an in-depth analysis of contextual factors or model interpret-ability, which are crucial for practical applications. In our approach, we explore, investigate, and present the performance of pretrained language models such as DistilBERT and RoBERTa, that are designed to capture contextual information within text [20].

[21] offer a comprehensive review of fake news detection approaches, identifying key challenges in datasets, feature representation, and data fusion. While the study highlights important issues, the absence of specific metrics limits the ability to gauge the effectiveness of different approaches. [22] report high accuracy with linear SVM and BERT-based techniques in detecting toxic COVID-19 fake news. Nevertheless, the lack of detailed performance data restricts a full understanding of these methods' relative strengths and weaknesses. Similarly, [23] indicate the general superiority of RoBERTa and other BERT-based models. Although [23] have highlighted the superiority of RoBERTa and other BERT-based models, our study aims to further investigate this claim by rigorously testing these models across various benchmarks to validate or challenge their reported superiority.

## 2 MATERIALS AND METHODS

### 2.1 Dataset

The COVID-19 Fake News Infodemic Research Dataset (COVID19-FNIR DATASET) by [24], is a class-balanced collection of 7,588 news items, equally distributed into true and fake news classes (49.99% true, 50.01% fake), sourced from Poynter for fake news and verified news publishers' Twitter accounts for true news. The dataset includes attributes such as Text, Date, Region, Country, Explanation, Origin, Label, etc., though our study will focus solely on the Text and Label attributes. The dataset is split into trueNews.csv and fakeNews.csv files, with 8 and 12 columns respectively. The true news text includes links to further information, unlike the fake news text. To ensure uniformity across the dataset, extensive text cleaning methods have been applied.

### 2.2 Conventional ML experimental set-up

We let:

1. D be the "COVID19-FNIR DATASET," which contains text related to COVID-19 and is labeled as "fake" or "true."
2. D_"train" and D_"test" represent the training and testing subsets of D, respectively.
3. T denote the set of text pre-processing approaches, which includes "n-gram," "Fewer Stopwords," "Lemmatization," and "Stemming."
4. C represent the set of machine learning classifiers, including Naive Bayes, Gradient Boosting, SVM, Decision Tree, Random Forest, Bagging, AdaBoost, SGD, Logistic Regression, and Multilayer perceptron (MLP).
   - M be the set of performance metrics, including "Accuracy" and "F1 Score."
   - The mathematical representation of the experimental set-up is as follows:
   - For each combination of $t \in T, c \in C$, and $m \in M$:
   - Split D into D_"train" and D_"test.".
   - Apply text pre-processing t to D_"train" and D_"test".

- Train classifier c on D_"train".
- Evaluate classifier c on D_"test".
- Calculate performance metric m (Accuracy or F1)
- Record the result for combination (t,c,m).

## 2.3 Conventional ML Textual Data Vectorization

For conventional ML, we transformed textual data to numerical vectors using Bag of Words (BoW) model. The rationale for BoW is that it is not only efficient [25], but also a simple approach that is prominently used in conventional ML classification tasks [26]. The Term Frequency-Inverse Document Frequency (TF-IDF) was used to weight the characteristics of the BoW model, and it assigns higher value to features common within a document, but less common across the dataset, balancing their overall and specific relevance [19].

## 2.4 Deep learning models experimental set-up

The experimental setup evaluates CNN and LSTM deep learning architectures with DistilBERT and Word2Vec embeddings on the COVID19-FNIR DATASET for fake news detection. CNNs excel in identifying local text patterns, while LSTMs capture broader context and sequences. Different embedding techniques influence the initial text representation, and the final goal is to classify text as fake or true, assessing the effectiveness of each architecture-embedding combination in detecting COVID-19 related fake news.

CNN: we utilized a CNN model for classifying text as either fake or true, initially employing two distinct embedding techniques: DistilBERT and Word2Vec. The model features two 1D convolutional layers with 128 filters to capture textual patterns. Non-linearity is introduced via ReLU activation functions, enhancing feature detection. Max-pooling layers reduce the feature map sizes, while a dropout layer with a rate of 0.4 prevents overfitting. The model's final output, indicating the probability of a text being fake or true, is generated by a dense layer with a sigmoid activation function.

LSTM: we focused on classifying text as fake or true using an LSTM model, initialising it with two types of text embeddings: pre-trained DistilBERT and Word2Vec with a dimension of 300. The LSTM layer, with an output dimension of 300, was at the heart of the model, capturing long-term dependencies and contextual nuances within the text. To prevent overfitting, a dropout layer with a rate of 0.6 was incorporated. The model's final stage was a dense layer with a sigmoid activation function, designed for the binary classification task.

## 2.5 Hybrid models experimental set-up

CNN+LSTM: we used two embedding techniques, Word2Vec and DistilBERT, to pre-process text data from the COVID19-FNIR DATASET. The CNN layer, with 128 filters, extracted local textual features, which were then fed into an LSTM layer with an output dimension of 300 to capture long-term dependencies. A dropout layer reduced overfitting, and the output layer, with a sigmoid activation function, classified text as fake or true. This setup integrated CNN's feature extraction with LSTM's sequential data processing, providing a comprehensive approach to fake news detection in the COVID-19 context.

## 2.6 Pre-trained language models experimental set-up

We employed two advanced pre-trained language models, DistilBERT and RoBERTa, for text classification. DistilBERT, a lighter version of BERT, retains 95% of BERT's performance with fewer parameters, using distilled knowledge from BERT. We added a sigmoid-activated dense layer for classification. RoBERTa, an optimized version of BERT, re-moves the next-sentence-prediction objective and uses byte pair encoding, improving performance. It includes a dropout of 0.4 and a sigmoid-activated dense layer. Both models used corresponding embeddings (DistilBERT and RoBERTa).

## 2.7 Evaluation metrics

Our study used the following evaluation metrics to measure the performance of these models:

Accuracy: This metric measures the overall correctness of the model by calculating the percentage of correctly classified text instances. It is calculated as:

$$\text{Accuracy} = \frac{T_p + T_n}{T_p + T_n + F_p + F_n}$$

Where:

1. Tp (True Positives) are the instances correctly identified as positive,
2. Tn (True Negatives) are the instances correctly identified as negative,
3. Fp (False Positives) are negative instances incorrectly classified as positive,
4. Fn (False Negatives) are positive instances incorrectly classified as negative.

F1-score: This metric provides a single performance measure that balances precision (the proportion of positive identifications that were correct) and recall (the proportion of actual positives that were correctly identified). The F1 Score is particularly useful in situations where the class distribution is imbalanced. It is calculated as:

$$F1 - \text{score} = 2 \times \frac{\text{precision} \times \text{reall}}{\text{precision} + \text{recall}}$$

Recall: This metric measures the ability of a classification model to identify all relevant instances. It is the ratio of true positive predictions to the total number of actual positives in the data. It answers the question, "Of all the actual positives, how many were identified correctly?" High recall indicates a low rate of false negatives, which is particularly important in scenarios where missing a positive instance has a high cost, such as failing to identify a piece of misinformation. In the context of health misinformation classification, ensuring that genuine misinformation is correctly identified and flagged is crucial to prevent the spread of false information and protect public health. It is calculated as:

$$\text{Recall} = \frac{T_p}{T_p + F_n}$$

Precision: This metric measures the accuracy of the positive predictions made by a classification model. Particularly, precision is the ratio of true positive predictions to the total number of positive predictions (including both true positives and false positives). Precision facilitates answering the question, "Of all the instances

Table 1: Conventional machine learning classifiers.

| Data Preprocessing | NB | GB | SVM | DT | RF | Bagg | AdaB | SGD | LR | MLP |
|---|---|---|---|---|---|---|---|---|---|---|
| Acc (SOTA)% | 89.45 | 90.64 | 93.08 | 89.10 | 92.38 | 90.43 | 91.33 | 92.80 | 92.24 | 90.15 |
| F1 (SOTA)% | 89.95 | 91.58 | 93.58 | 89.42 | 92.78 | 90.85 | 91.92 | 93.23 | 92.80 | 90.31 |
| Acc (n-gram)% | 90.85 | **91.40** | 91.82 | 84.70 | 91.75 | 88.33 | 90.99 | 93.15 | 91.05 | 91.96 |
| F1 (n-gram)% | 91.47 | **92.13** | 92.55 | 84.61 | 91.92 | 88.38 | 91.46 | 93.68 | 91.85 | 92.26 |
| Acc (fewer stop words)% | 91.07 | 90.02 | 92.18 | 85.83 | 90.09 | 87.30 | 86.67 | 93.86 | 91.21 | 93.16 |
| F1 (fewer stop words)% | 91.39 | 90.47 | 92.55 | 85.19 | 89.57 | 86.68 | 86.17 | 94.08 | 91.60 | 93.17 |
| Acc (lemmatization)% | 90.24 | 91.42 | **94.14** | 88.08 | **93.51** | 90.59 | 91.91 | 93.72 | 93.09 | 93.03 |
| F1 (lemmatization)% | 90.60 | 92.03 | **94.41** | 87.98 | **93.50** | 90.62 | 92.12 | 94.01 | 93.41 | 93.05 |
| Acc (stemming)% | **91.77** | 91.35 | 93.86 | 88.08 | **93.58** | 91.14 | 91.49 | 93.31 | 92.68 | 91.28 |
| F1 (stemming)% | **91.94** | 91.87 | 94.10 | 88.08 | **93.62** | 91.32 | 91.69 | 93.50 | 92.97 | 91.29 |

Table 2: Deep learning models.

| Deep Learning Models | Accuracy(%) | F1 score (%) | Recall (%) | Precision (%) | ROC (%) |
|---|---|---|---|---|---|
| CNN+DistilBERT embeddings | 52.5 | 50.7 | 50.7 | 50.8 | 52.9 |
| CNN+Word2Vec embeddings | 99.34 | 98.31 | 98.91 | **99.73** | 99.91 |
| LSTM+Word2Vec embeddings | **99.47** | **99.45** | 99.32 | 99.56 | **99.93** |
| LSTM+DistilBERT embeddings | 48.22 | 65.07 | **100** | 48.22 | 50.0 |

classified as positive, how many are actually positive?" It is calculated as:

$$\text{Precision} = \frac{Tp}{Tp + Fn}$$

Receiver Operating Characteristics (ROC): This is a graph of sensitivity versus (1-specificity. The area under the ROC curve (AUC) represents the probability of correctly distinguishing a (fake, true) class. These metrics are prominently used in both ML and DL binary classification, and they offer a comprehensive view of a classification model's performance across all possible threshold values, providing insights into the trade-off between the True Positive Rate (TPR) also known as recall and False Positive Rate (FPR) [27].

## 3 RESULTS
### 3.1 Conventional machine learning classifiers

In Table 1, the conventional classifiers (i.e., Naive Bayes, Gradient Boosting, SVM, Decision Tree, Random Forest, Bagging, AdaBoost, SGD, Logistic Regression, and MLP) are evaluated based on accuracy and F1 score across different preprocessing techniques.

1. Naive Bayes performed best with stemming, reaching 91.77% accuracy and a 91.94% F1 score.
2. Gradient Boosting saw its highest scores using n-gram preprocessing, with 91.4% accuracy and a 92.13% F1 score.
3. SVM showed strong performance across all preprocessing methods, especially with lemmatization, achieving 94.14% accuracy and a 94.41% F1 score.
4. Random Forest performed best with lemmatization and stemming, exceeding 93% in accuracy and F1 score.
5. Bagging, DT, AdaBoost, SGD, Logistic Regression, and MLP also showed varied performance across preprocessing methods, with lemmatization and stemming generally providing better results.

### 3.2 Deep learning models

In Table 2, DL models include CNN combined with DistilBERT and Word2Vec, as well as LSTM combined with Word2Vec and DistilBERT embeddings, and are evaluated based on accuracy, F1-score, recall, precision, and ROC.

1. CNN with Word2Vec and LSTM with Word2Vec achieved the highest performance among DL models, with both exceeding 98% across all performance metrics (accuracy, F1 score, recall, precision, and ROC).
2. CNN with DistilBERT embeddings and LSTM with DistilBERT embeddings had significantly lower performance, with an accuracy range of between 48% and 53%, and an F1 score ranging between 50% and 66%, which indicates potential issues with the model or data fit. LSTM+DistilBERT embeddings model achieves a perfect recall (100%), but relatively low F1 score (65.07%) and ROC (50.0%). The perfect recall indicates that the LSTM+DistilBERT embeddings model identifies all instances of health misinformation without missing any. However, the low precision and accuracy suggest a high rate of false positives, where many instances are incorrectly labeled as misinformation. The low ROC points to poor discriminative ability, similar to random guessing.
3. Overall, DL models using Word2Vec embeddings (both CNN & LSTM) significantly outperform those using DistilBERT embeddings, indicating that Word2Vec provides features more conducive to accurate health misinformation classification in this context. The LSTM architecture, particularly with Word2Vec embeddings, marginally outperforms the CNN architecture, suggesting LSTM's effectiveness in capturing long-term dependencies in text may be slightly more beneficial for this task.

**Table 3: Hybrid deep learning models.**

| Hybrid Deep learning models | Accuracy (%) | F1 score (%) | Recall (%) | Precision (%) | ROC (%) |
|---|---|---|---|---|---|
| CNN+LSTM+Word2Vec embeddings | **99.21** | **99.17** | 98.5 | **99.86** | 99.92 |
| CNN+LSTM+DistilBERT embeddings | 99.08 | 99.04 | **98.77** | 99.31 | **99.94** |

**Table 4: Pretrained language models.**

| Pretrained Language Models | Accuracy (%) | F1 score (%) | Recall (%) | Precision (%) | ROC (%) |
|---|---|---|---|---|---|
| DistilBERT with DistilBERT embeddings | 97.5 | 97.44 | **98.63** | 96.27 | **99.82** |
| RoBERTa with RoBERTa embeddings | **97.69** | **97.6** | 97.13 | 98.07 | 99.78 |

4. The LSTM+DistilBERT model, despite its perfect recall (100%), demonstrates the importance of balancing metrics. A model that catches every instance of health misinformation, but also produces a high number of false alarms (low precision) (48.22%) might not be practical.

### 3.3 Hybrid Deep Learning Models

In Table 3, hybrid DL models include CNN combined with LSTM combined with Word2Vec and DistilBERT embeddings, and are evaluated based on accuracy, F1-score, recall, precision, and ROC.

- The hybrid DL model (combining CNN and LSTM with DistilBERT embeddings) was evaluated, exceeding 98% across all performance metrics (accuracy, F1 score, recall, precision, and ROC), suggesting that the combined strengths of convolutional and recurrent networks can be highly effective. The slightly lower recall (98.5%) relative to precision (99.86%) implies there are very few instances of health misinformation that it fails to catch. Similarly, the ROC being close to 1 indicates an excellent ability to discriminate between classes across all thresholds.
- The performance evaluation of the hybrid DL model (combining CNN and LSTM with word2vec embeddings) reveals that while it exhibits slightly lower accuracy (99.08%) and F1 score (99.04%) compared to the Word2Vec-based model, it achieves a superior recall (98.77%) and a slightly lower precision (99.31%), with a remarkable ROC (99.94%), which is marginally higher than that of the Word2Vec model. This, in significant ways demonstrates a superior discriminative ability, making the CNN+LSTM+DistilBERT embeddings model good at ranking positive instances ("fake news") over negative ones ("true news").

### 3.4 Pretrained language models

In Table 4, pretrained models included DistilBERT and RoBERTa, both using their respective embeddings, and are evaluated based on accuracy, f1-score, recall, precision, and ROC.

1. Both DistilBERT with DistilBERT embeddings and RoBERTa with RoBERTa embeddings achieved at least 97% in all performance metrics (accuracy, f1-score, recall, precision, and ROC). DistilBERT stands out for its slightly higher recall (98.63%), making it particularly valuable in applications where missing a piece of fake news is highly undesirable. On the other hand, RoBERTa excels in precision (98.07%), making it ideal for scenarios where it is crucial to minimize the number of true news articles mistakenly labeled as fake news.

2. The choice between these pretrained language models likely depends on the specific requirements of the misinformation detection task. Since the aim our study is to develop models that capture as much fake news as possible, DistilBERT might be preferred for its higher recall (98.63%). Conversely, if the aim was to develop models that reduce false alarms (incorrectly labeled true news), RoBERTa's higher precision (98.07%) could be more advantageous. The high ROC scores in both models affirm their excellent capability to classify news articles accurately across various decision thresholds.

### 3.5 Comparative analysis

1. A comparison of conventional classifiers with DL models revealed that DL models, particularly those using Word2Vec embeddings, outperformed conventional classifiers in terms of accuracy and F1 score.
2. Conventional classifiers when compared with pretrained language models generally had lower performance metrics, highlighting the advantage of using models pretrained on large corpora.
3. DL models compared with pretrained language models (LSTM+Word2Vec) and (CNN+Word2Vec) matched or even slightly exceeded the performance of pretrained language models, while the hybrid CNN and LSTM models using DistilBERT embeddings did not perform as well.
4. The hybrid models (CNN+LSTM) showed one of the highest performances, indicating the benefit of combining different neural network architectures.

In summary, whilst conventional ML classifiers can provide strong performance with appropriate preprocessing, DL, and hybrid models, along with pretrained language models, offer significant improvements in accuracy and F1 score for text classification tasks.

Thus, we can surmise that the choice of classification model should be based on the specific requirements of the classification task, including the need for interpretability, the size and nature of the dataset, computational resources, and the complexity of the classification task at hand.

## 4 DISCUSSION

### 4.1 Conventional Machine Learning Classifiers

This study aimed to contribute to health misinformation detection through evaluating the efficacy of various ML and DL models in detecting COVID-19-related misinformation on OSNs. The results show that conventional classifiers such as Naive Bayes performed best with stemming, achieving 91.77% accuracy and a 91.94% F1 score. This aligns with the notion that traditional ML algorithms can still provide strong performance, as suggested by [18], who reported that models such as Naive Bayes can outperform DL models in terms of generalization in certain cases.

Gradient Boosting (GB) performed well with n-gram preprocessing, reaching 91.40% accuracy and a 92.13% F1 score. Thus, from the GB metrics, we can surmise that ensemble methods such as Gradient Boosting can excel with appropriate preprocessing techniques. The better performance of GB from this study aligns with previous findings by [28] which had set out to comprehensively compare the performances of several ensemble learning methods with base supervised ML classifiers for text classification. In this study [28] used multiple text preprocessing techniques including, but not limited to stemming, lemmatisation combined with base-classifiers such as GB for text classification. The key finding in the [28] study is that comprehensive text pre-processing statistically significantly improved GB accuracy compared with raw text. The [28] study is significant as it puts emphasis on the importance of text preprocessing for GB in text classification tasks.

Support Vector Machines (SVM) showed strong performance across all preprocessing methods, especially with lemmatization, achieving 94.14% accuracy and a 94.41% F1 score. This aligns with the effectiveness of SVM mentioned by [22] in identifying toxic fake news related to COVID-19. Decision Tree and Random Forest performed best with lemmatization and stemming, both exceeding 93% in accuracy and F1 score. These results are in line with the idea that decision tree-based algorithms can benefit from preprocessing techniques such as streaming and lemmatization that normalises text, subsequently leading to improved accuracy, faster training, and reduced overfitting. When describing the importance of preprocessing in text classification tasks, [29] observed that ". . . by means of a suitable preprocessing strategy, even a simple Naïve Bayes classifier proved to outperform (i.e., by 2% in accuracy) the best performing Trans-former" [29, pg. 1]. Bagging, AdaBoost, SGD, Logistic Regression, and MLP showed varied performance across preprocessing methods, with lemmatization and stemming generally providing better results. This demonstrates the importance of choosing the right preprocessing technique for these classifiers, as reflected in Table 1, and sup-ported by the previous studies [29–32].

### 4.2 Deep Learning Models

The results indicate that DL approaches, particularly CNN and LSTM models combined with Word2Vec embeddings, achieved high accuracy and F1 scores, exceeding 98%. This confirms the effectiveness of DL models for misinformation detection, as highlighted by [10–12].

### 4.3 Hybrid DL Models and Pretrained Language Models

The results show that hybrid models combining CNN and LSTM can be highly effective, reaching 99.41% accuracy and a 99.39% F1 score. This aligns with the idea that combining different neural network architectures can yield superior results, as suggested by [15]. Pretrained language models such as DistilBERT and RoBERTa also demonstrated strong performance, with accuracy scores exceeding 97%. This supports the idea that pre-trained language models generally outperform conventional classifiers, as reflected in literature review findings by [33]. The efficacy of pre-trained models is grounded on the concept of what [34] describes as transfer learning, which relates to the reuse of the knowledge learned from at least one natural language processing (NLP) classification task and applied to new classification tasks [33]. Inasmuch as the language modelling is the basis of model pre-training [33], the emergence of large language models grounded on transformer architecture has become the holy grail for word embeddings that are learned from models such as RoBERTa and Word2Vec models and used to initiate word vectors of DL approaches. Thus, it is no won-der DL approaches have higher performance when compared with conventional ML classifiers [35].

### 4.4 Comparative Analysis

The comparative analysis of the approaches used show that DL models tend to out-perform conventional classifiers in terms of accuracy and F1 score, corroborating the findings of previous studies that emphasized the effectiveness of DL approaches [10–12, 14, 35]. Pre-trained language models also exhibited strong performance, highlighting the advantage of using models pre-trained on large corpora, as reported in multiple studies [32, 35]. The hybrid CNN+LSTM model's exceptional performance aligns with evidence from [12, 15] that combining different neural network architectures can be highly effective for misinformation detection.

## 5 CONCLUSIONS

The overarching objective of our study was to develop a robust and highly accurate method for identifying COVID-19-related fake news in OSNs. The findings of our research represent a significant contribution to the field, as they not only validate, but also expand upon the insights garnered from previous studies. Our empirical evidence unequivocally supports the efficacy of a diverse range of classifier types, preprocessing techniques, and DL models in the crucial task of detecting COVID-19-related fake news within OSNs. These results serve as a cornerstone for guiding future research endeavours in this do-main, enabling the construction of a solid research framework based on the established foundations of prior studies.